\documentclass[sigconf,authorversion,nonacm]{acmart}

\AtBeginDocument{\providecommand\BibTeX{{\normalfont B\kern-0.5em{\scshape i\kern-0.25em b}\kern-0.8em\TeX}}}

\setcopyright{none}

\newcommand{\Resources}{\textit{Resources:\ }}
\newcommand{\LevelDesign}{\textit{Level Design:\ }}
\newcommand{\Instructions}{\textit{Instructions:\ }}
\newcommand{\Concept}[1]{\medskip\noindent\textbf{#1}}

\begin{document}

\title[One Pixel, One Interaction, One Game]{One Pixel, One Interaction, One Game}
\subtitle{An Experiment in Minimalist Game Design}

\author{Pier Luca Lanzi}
\authornote{Contact author.}
\email{pierluca.lanzi@polimi.it}
\orcid{0000-0002-1933-7717}
\affiliation{\institution{Politecnico di Milano}
	\city{Milano}
	\country{Italy}
}

\author{Daniele Loiacono}
\email{daniele.loiacono@polimi.it}
\orcid{0000-0002-5355-0634}
\affiliation{\institution{Politecnico di Milano}
	\city{Milano}
	\country{Italy}
}

\author{Alberto Arosio}
\affiliation{\institution{Digital Tales}
	\city{Milano}
	\country{Italy}
}

\author{Dorian Bucur}
\affiliation{\institution{Universit\'a degli Studi di Milano}
	\city{Milano}
	\country{Italy}
}

\author{Davide Caio}
\affiliation{\institution{CyberCoconut}
	\city{Milano}
	\country{Italy}
}

\author{Luca Capecchi}
\affiliation{\institution{Politecnico di Milano}
	\city{Milano}
	\country{Italy}
}

\author{Maria Giulietta Cappelletti}
\affiliation{\institution{Politecnico di Milano}
	\city{Milano}
	\country{Italy}
}

\author{Lorenzo Carnaghi}
\affiliation{\institution{Politecnico di Milano}
	\city{Milano}
	\country{Italy}
}

\author{Marco Giuseppe Caruso}
\affiliation{\institution{Politecnico di Milano}
	\city{Milano}
	\country{Italy}
}

\author{Valerio Ceraudo}
\affiliation{\institution{Politecnico di Milano}
	\city{Milano}
	\country{Italy}
}

\author{Luca Contato}
\affiliation{\institution{Rising Pixel}
	\city{Gran Canaria, Canary Islands}
	\country{Spain}
}

\author{Luca Cornaggia}
\affiliation{\institution{Politecnico di Milano}
	\city{Milano}
	\country{Italy}
}

\author{Christian Costanza}
\affiliation{\institution{Big Bang Pixel}
	\city{Milano}
	\country{Italy}
}

\author{Tommaso Grilli}
\affiliation{\institution{Politecnico di Milano}
	\city{Milano}
	\country{Italy}
}

\author{Sumero Lira}
\affiliation{\institution{Neotenia}
	\city{Milano}
	\country{Italy}
}

\author{Luca Marchetti}
\affiliation{\institution{Studio Evil}
	\city{Bologna}
	\country{Italy}
}

\author{Giulia Olivares}
\affiliation{\institution{Lola Slug}
	\city{Milano}
	\country{Italy}
}

\author{Barbara Pagano}
\affiliation{\institution{Universit\'a degli Studi di Milano}
	\city{Milano}
	\country{Italy}
}

\author{Davide Pons}
\affiliation{\institution{Politecnico di Milano}
	\city{Milano}
	\country{Italy}
}

\author{Michele Pirovano}
\affiliation{\institution{Curiosity Killed the Cat}
	\city{Bergamo}
	\country{Italy}
}

\author{Valentina Tosto}
\affiliation{\institution{AnotheReality}
	\city{Milano}
	\country{Italy}
}

\renewcommand{\shortauthors}{Lanzi and Loiacono, et al.}

\ccsdesc[500]{Applied computing~Computer games}
\ccsdesc[300]{Human-centered computing~Interaction devices}
\keywords{game design, game mechanics, player interaction, player experience}

\begin{abstract}

Minimalist game design was introduced a decade ago as a general design principle with a list of key properties for minimalist games: basic controls, simple but aesthetically pleasing visuals, interesting player choices with vast possibility spaces, sounds that resonate with the design. In this paper, we present an experiment we did to explore minimalism in game using a bottom-up approach. We invited a small group of professional game designers and a larger group of game design students to participate in a seminal experiment on minimalism in game design. 
We started from the most basic game elements: one pixel and one key which provide the least amount of information we can display and reasonably the most elementary action players can perform. 
We designed a game that starts with a black pixel and asks players to press a key when the pixel turns white. This minimal game, almost a Skinner box, captures the essential elements of the mechanics of games like “The Impossible Game,” which asks players to do nothing more than press a key at the right moment. We presented this game concept to the professional game designers and challenged them to create other games with the least amount of player interaction and displayed information. We did not specify any constraint (as usually done in other contexts) and left them free to express their view of minimalistic game design. We repeated the experiment with 100+ students attending a master-level course on video game design and development at our institution. We then analyzed the creations of the two groups, discussing the idea of minimalistic design that emerges from the submitted game concepts. \end{abstract}

\maketitle

\section{Introduction}
\label{sec:introduction}
Minimalism in games comes in a wide variety of forms \cite{DBLP:conf/digra/Juul07}. In tabletop games, it may be achieved through abstraction \cite{wiki:AbstractStrategyGames}, the complexity of rulesets \cite{Breakthrough}, or a visual redesign, as in the recent revamp of Uno decks \cite{wiki:Uno,MinimalistUno}.  In video games, minimalism can be introduced at visual level, at control level or at system level \cite{DBLP:conf/digra/Myers09,DBLP:conf/fdg/NealenSB11}. 
Minimalist video games can be visually abstract and based on a  stylized representations of a reference game world \cite{140,ThomasWasAlone}; often, visual abstraction is accompanied with minimal sound design \cite{DarkEcho}. They may have simplified controls \cite{OneButton} that map into complex behaviors \cite{Canabalt,Downwell}. They may be based on minimalist systems providing players with a limited number of interesting choices that enable surprisingly deep gameplay. A decade ago, Nealen et al. \cite{DBLP:conf/digra/Myers09,DBLP:conf/fdg/NealenSB11} introduced the notion of \textit{Minimalist Game Design},  discussing the importance of self-imposed, artificially designed constrains for exploring new frontiers of gameplay. 
Over the last decade, the principles of minimalist game design have been embraced by the increasing number of game jams, organized all over the world,
	which are inherently based on self-imposed, deliberate constrains \cite{LowrezJam,GameboyJam}, a major enabler of creativity in design 	\cite{DBLP:conf/fdg/NealenSB11}.

In this paper, we analyze minimalist game design using a bottom up approach. We did not surveyed existing games as \cite{DBLP:conf/fdg/NealenSB11}, instead we challenged professional and video game design students to develop the most minimalist games they could think of and analyze their creations. 
Our goal was twofold. We wanted to explore minimalism in game design a decade after \cite{DBLP:conf/digra/Myers09,DBLP:conf/fdg/NealenSB11} both from the perspective of professionals working in the industry and students who are approaching the field. Furthermore, we wanted to create a collection of examples of minimalist games that we could provide to future students as inspiration (a sort of minimalist game design sandbox to play with) for developing their own idea of minimalism.
We designed the most basic game we could come up with (involving just a pixel and an elementary interaction) and explored level design in such extreme minimalist scenario.
Initially, we invited a small group of professional game designers to participate in the experiment, receiving more than 20 concepts. We did not enforce constraints as it usually happens during game jams. We just provided them the description of our very basic game as inspiration to design other original games with the least amount of player interaction and on-screen information. We asked them to submit a brief  concept describing the gameplay, the level design principles, and the instructions to teach players how to play. Next, we repeated the experiment by inviting the 100+ students attending the master-level video game design and development course at our institution, receiving 100+ concepts. Finally, we analyzed the submitted concepts and investigated the idea of minimalist game design that emerges in the two subject groups, also with respect to the seminal work of \cite{DBLP:conf/digra/Myers09,DBLP:conf/fdg/NealenSB11}.

\section{Background}
\label{sec:related}
Minimalism is an artistic movement that originated in the late 1950s and developed throughout the 1960s.  It is characterized by extreme austerity and simplicity that seek to uncover the essence of a subject by eliminating all non-essential parts of it. Initially primarily focused on music and aesthetics, it has expanded its influence over all sorts of media, and human activities \cite{MarieKondo}, including videogames \cite{DBLP:conf/digra/Myers09,DBLP:conf/fdg/NealenSB11}.

Videogames were born minimal out of necessity to deal with the severe limitations of early hardware platforms \cite{RacingTheBeam}. As the technological resources were becoming practically limitless, minimalism helped small independent companies deal with other, still existing, time and labor limitations \cite{DBLP:conf/fdg/NealenSB11}, becoming a distinctive stylistic reference for many indie games (e.g., \cite{SuperHexagon,TheImpossibleGame}). 
Minimalism is usually associated with an idea of austerity, spareness, and simplicity. In videogames, it has been used as a tool to search for the essence of game definitions \cite{ DBLP:conf/digra/Myers09, DBLP:conf/digra/Juul03} and proposed as a general design principle \cite{DBLP:conf/fdg/NealenSB11,DBLP:journals/tciaig/IsaksenGTN18}. 

Juul \cite{DBLP:conf/digra/Juul03} analyzed seven game definitions from a variety of historical and contemporary sources. His approach was inclusive \cite{DBLP:conf/digra/Myers09} and aimed at proposing a definition that consolidated well-established and accepted knowledge. Juul \cite{DBLP:conf/digra/Juul03} synthetized a list six elements that are “necessary and sufficient for something to be a game”: (i) rules, (ii) variable and quantitative outcomes, (iii) valorization of outcome, (iv) player effort, (v) player attachment to outcome, and (vi) negotiable consequences. In contrast, Myers \cite{DBLP:conf/digra/Myers09} took an exclusive approach and eliminated what is not equally shared by all definitions. The result is a minimal list of four elements that defined games, (i) ``prohibitive'' rules or rules of denial, (ii) goals, including the game's winning conditions, (ii) opposition, provided for example by an antagonist, and (iv) representation, or a falseness that is contrary to the real. The list is presented as a tool to identify what should not considered a game. For example, Myers \cite{DBLP:conf/digra/Myers09} argues that  crossword puzzles are not games since they do not have element of opposition or an antagonist. 

Nealen et al. \cite{DBLP:conf/fdg/NealenSB11} proposed \textit{Minimalist Game Design} as a general design principle to elicit the exploration of new directions in design and gameplay by introducing self-imposed, deliberate constraints on the entire design and development process. Minimalist games should be based on essential mechanics, small rulesets, narrow decision spaces, abstract aesthetics; at the same time, the imposed constraints however should not limit the depth of play or the possibility space \cite{DBLP:conf/fdg/NealenSB11}. 

Minimalist games are also laboratories to study the characteristics of games in a smaller and more controlled environment. Iraksen et al. \cite{DBLP:journals/tciaig/IsaksenGTN18} proposed a toolset of techniques to evaluate the difficulty of game variants, to balance games, to explore the game space, and predict the likelihood of player achieving specific final scores. Their approach integrated several methods (e.g., automatic playtesting, Monte Carlo simulation, player modeling, and survival analysis) and would have been difficult to validate using more complex games \cite{DBLP:journals/tciaig/IsaksenGTN18}. Accordingly, 
Iraksen et al. \cite{DBLP:journals/tciaig/IsaksenGTN18} used a parametrized version of the minimalist game, Flappy Bird \cite{FlappyBird}, as an experimental laboratory to assess their methodology. Lankes \cite{DBLP:conf/chiplay/Lankes20} used  abstract visuals and gaze-based interaction to investigate the perceived quality of social communication towards abstract non-player characters using a minimalist game with basic mechanics.

\section{One Pixel, One Interaction, One Game}
\label{sec:one_pixel}

A white pixel is the minimum amount of information we can show on-screen, and pressing a key (or a button) is the least interaction we can ask players. Therefore, we can design a game where the player must press a key when the pixel turns white. The design of a level depends on the rate at which the pixel lights up and how long the pixel stays illuminated. For example, we can design a level where the pixel turns off for one second and on for one second; the next level could reduce the time the pixel stays on or when the pixel is off. Alternatively, we could have a level where the time the pixel stays on decreases as the level progresses; or a level with no pattern whatsoever, in which the switch-off and switch-on times are entirely random.
Such a game is elementary, but from a mechanical point of view, it captures essential components of games like ``The Impossible Game" \cite{TheImpossibleGame}, which asks players to do nothing more than to press a key to jump at the right time. We could change the interaction pattern by simply asking the player to hold down a key when the pixel is off and release it when the pixel is on, in a sort of negative space of the original interaction model. ``Fotonica'' \cite{Fotonica} has the same interaction model in which players hold the key down to run, release it to jump, and press it again to land. The change in the interaction is small but holding the spacebar to run and releasing it to jump makes the gameplay feel pretty different. We could replace the pixel states with sounds and ask the players to press a key when they hear specific sounds or music and ask players to press the key following the underlying rhythm. Similarly, ``The Impossible Game'' uses rhythm to help players understand when to jump. Following the same principle, we can create a small memory game: the white pixel lights up a given number of times (for example, three times in sequence), and then players must press the key for the same number of times.  
\section{An Experiment in Minimalist Game Design}
\label{sec:experiment}
The literature exploring minimalist game design takes a top-down approach and either propose it as a general design principle \cite{DBLP:conf/fdg/NealenSB11,DBLP:journals/tciaig/IsaksenGTN18} or synthesizes what are the core features of minimalist games \cite{DBLP:conf/digra/Juul03}. In our study, we decided to take a bottom-up approach and investigate what people already working in (or just approaching) the field view as minimalist game design. Accordingly, we contacted a group of professional game designers and challenged them to develop the most minimalist games they could think of and analyze their creations. We did not enforce any constraints as it usually happens during game jams. We just tried to inspire them with some minimalist games (the same described in Section~\ref{sec:one_pixel}) and asked them to design other original games with the least amount of player interaction and on-screen information. We asked them to submit a brief concept describing the gameplay, the level design principles, and the instructions to teach players how to play. Next, we invited the 100+ students attending a master-level video game design and development course at our institution by providing the same information and form provided to the first group of subjects. Overall, we received 22 concepts from professional designers and 104 from the master students.

\section{Statistics}
\label{sec:statistics}
We analyzed the submitted concepts and labeled them based on their genre. In particular, given the limited number of concepts and the high variety of ideas, we used high-level labels such as, Action, Memory, Puzzle, Rhythm, Exploration, and Party. We also label the games based on the skills they required (e.g., dexterity, reaction time, coordination, memory), and whether 
(i)    they were single or multi player;
(ii)   they had a reactive or stateful gameplay, that is, players' actions could be based solely by the current game state or also had to take into account previous states;
(iii)  they involved some strategic thinking;
(iv)   they were organized as one infinite level or as a sequence of levels, or
(v)    they applied procedural content generation. 
In terms of genre, 63\% of the proposed concepts were action games, 15\% were memory games, 11\% puzzles, 7\% rhythm games. In terms of players' abilities, 30\% involved some kind of dexterity, 27\% quick reaction time, 23\% some cognitive ability, 17\% memory, 8\% involved story telling. Most games were single players (93\%) and only very few games (3\%) required strategic decision. 62\% of the games were purely reactive, in that players did not need to remember the effect of previous actions; whereas, 38\% were stateful and asked the player to be aware of how the play developed over time. In terms of level design, 55\% were based on a single level that would run until gameover, while the remaining ones were organized as a series of separate levels; 80\% of the games applied procedural content generation.

\section{The Games}
\label{sec:games}
The submitted game concepts offered a wide variety of mechanics and themes. In this section, we include a small subset of them that
	we selected to provide a good overview of such a variety.
They are listed anonymously in alphabetical order to avoid any sort of bias. 

\Concept{3.. 2.. 1.. Go!}
Players have to fly an aircraft, represented with a pixel, collecting pixels that appear on the screen over time. The pixel aircraft starts to the center-left side of the screen, following a linear trajectory to the right side of the screen. If the player does not press a key in around 0.75 seconds (corresponding to a rhythm of 80bpm, or beats per minute), the aircraft starts following a broad counterclockwise spiraling trajectory that will reduce its radius over time until a three pixels radius is reached. Then the aircraft will start following a clockwise spiral of increasing radius. Players control the aircraft by pressing and releasing the key rhythmically with a frequency of at least 80bpm, playing a sound at each keystroke. The world is a toroid; when the aircraft exit the screen from one side, it will re-enter from the opposite side, maintaining the same trajectory. The aircraft starts with four fuel units, and this will drop by one unit every ten seconds; when it reaches zero, the game ends. Players must harvest the fuel pixels that appear on the screen and remain there for a limited time; when the time is up, the fuel pixel will start flashing before it disappears and another pixel appears. The aircraft is also the game user interface. A short flash every two seconds means that the aircraft has three fuel points; two flashes indicate two remaining fuel points; three flashes warn players that just a single fuel unit remains. When the aircraft has collected the last fuel pixel, it starts to flash at a high frequency while playing a piece of music composed of three sounds, and it moves to a new starting position. \Resources three sounds, two pixels, one key. \LevelDesign the game is a rhythm game, and each level is carefully designed to play (if performed perfectly) to a given melody (or rather a rhythm). Every level has a growing difficulty represented by the pixels' time to collect that decreases with the next level, making it more difficult to maintain the fuel. \Instructions Fearless pilot! Press a button to turn the plane clockwise, release it to turn counterclockwise. Refuel by collecting ten pixels and move to the next level. Let the pace be with you! 

\Concept{Aliens Attack from Deep Space.}
When the output system announces an alien presence (on a screen or something else), the player has little time to shoot and kill it. 
\Resources An input system (a button, better if a trigger on a gun) and an output system (a pixel that turns on, a sound, a vibration).
\LevelDesign aliens will arrive with a random time ranging from 1 second to one minute, such high variability aims at increasing players anxiety over the waiting; initially, two aliens may arrive just a second away with 33\% probability; then, arrivals will be completely at random; after a given time, more than one shot might be required to kill aliens in later levels (two shots at level 2, three shots at level 3, etc.); random cool down periods between levels will increase anxiety and puzzle players.
\Instructions It is 1956 and you are the only person left on the face of the earth; you are in your bunker and watch the outside world from your peep hole; the aliens are attacking and your only chance is to kill them as soon as they are visible from your bunker peephole.

\Concept{Circle Wave.}
A circle appears and disappears at different positions on a touch screen. 
The player must keep her finger on the screen and must avoid the circle by moving his finger, without ever lifting it from the screen.
The circle slowly grows in size as the game progresses.
The game ends when either the player lifts the finger from the screen or it touches the circle.
\Resources a white circle and a touch screen. We could use two contrasting colors for the circle and the background. 
\LevelDesign Initially, the circle grows slowly, covering most of the screen in around three seconds and it shrinks until it disappears in around two seconds. As the game continues, the circle grows faster and shrinks faster. 
\Instructions Keep your finger on the screen and avoid the white circle. 

\Concept{Constellation.}
Constellation is a memory game inspired by the popular game "connect the dots".
White dots appear on the screen. When a dot is hovered (using a finger on a touch screen or a mouse pointer), another dot starts flashing.
When a dot it is no longer hovered, it starts to fade out slowly.
The last dot of the constellation activates the first dot of constellation.
Players must hover all the dots, following a predefined sequence, as quickly as possible so that when they reach the last dot, all the constellation is blinking at once. 
\Resources A display with two colors and an input system (a touch screen or a mouse).
\LevelDesign A constellation it's a loop of dot; level difficulty depends on the complexity of the shapes and the time before hovered dots disappear.
The faster dot that fades out faster, define the minimum time required to complete the level.
	The game should be a relaxed looping activity in which player firstly discover the dot positions, then repeat the path to activate the dots all together.
\Instructions Hover near to the next flashing dot to complete the constellation.  
\Concept{Dot-A-Mole.}
The player must captures the dots appearing on a touch screen before they turn off. When players touches a dot a sound is played.
\Resources a pixel (or a shape); two colors for the shape and the background; a sound to notify that the target has been hit; a touch screen. 
\LevelDesign a level consists of a given number of pixels; initially, the rhythm is slow and steady; as the game progresses, the rhythm increases and might become irregular while the time the pixel is active decreases; 
in each level, players must capture an increasing percentage of pixels. 
\Instructions Catch as many pixels as you can.

\Concept{Follow.}
A pixel moves on the screen in a pseudorandom way, moving away from the player's mouse pointer.
The player must try to stay as close as possible to the pixel.
\Resources a pixel and a mouse.
\LevelDesign player score is based on the distance between the mouse pointer and the pixel. As the score increases, the pixel moves faster. A game lasts for around two minutes. 
\Instructions stay as close as possible to the escaping pixel.

\Concept{Hop!}
The game is inspired to children jump rope games. Two identical squares are positioned at a short distance from each other. 
One, the avatar, is still; the other one, the obstacle, rotates around its center. 
Initially the avatar is positioned to the right of the obstacle. 
Players controls the avatar square using a key. Every time they hit the key, the avatar jumps in the direction determined by angle of the nearest side of the obstacle. So for instance, if players hit the key when the obstacle is at 90 degrees, the avatar will jump vertically since the nearest obstacle side is vertical; if players blink their eyes when the obstacle is rotate at 45 degrees, the avatar will jump with a 45 degrees direction to the left or to the right, depending on the obstacle position. 
The goal of the game is to jump back and forth from one side to the other side of the obstacle as many times as possible.
When players hit the obstacle, the game ends.
\Resources Two squares, one for the obstacle, one for the avatar. A palette of two colors. One key.
\LevelDesign Each level lasts for ten full obstacle rotations. Initially, the obstacle rotates very slowly and only in one direction.
Rotation speed slightly increases with the subsequent levels. Rotation direction might also change during one level, later in the game. 
\Instructions Press the key to jump the diamond as many times as you can.

\Concept{I am the Fastest!}
When the output system notifies the beginning of a match (for example, turning a pixel on or playing a sound), two players take turns and have to press a key (or push a button) as many times as possible before the output system notifies the end of the turn (turning the pixel off or playing another sound); a player's score is computed as the number of times the key was pressed during the match, minus the times it was pressed outside the turn (so that it is inconvenient for players to start pressing the key before and after their turn) the match ended. The player with the highest score wins.
\Resources An input system (a button will do the job) and an output system (a pixel that turns on, a played sound, a momentary vibration).
\LevelDesign There is no level design.
\Instructions Who is the fastest? Take turns with a friend. When light is on, push the button as many times as you can before the light turns off; be careful, pushing the button when the light is off will result in a penalty. At the end, the light will flash once, if the first player has won; twice if the second player has won; three times in case of a tie.

\Concept{Kick The Engine.}
All engineers know that to make things work sometimes the best solution to kick it. In this minimalist racing game, players must beat their car engine to make it go fast and overtake opponents.
But not too fast, otherwise players might lose the control of the car and be forced to stop. The screen shows the car position in the race using a line of 15 pixels (from first to fifthteen) 
	and a 4x4 pixels grid that represents the distance of next opponent to overtake and the finish line as a checkered flag.
Players speed up by pressing a key (or tapping on an touch screen), the speed decreases if players do not interact; 
players must keep an adequate speed to overtake opponent while keeping the control of the car.
The grid gives players an indication of how fast they should go to be able to overtake the opponent car: if the square decreases in size, players should speed up; if the square increases in size, players are getting near the opponent car and should be careful not too speed up to much and lose control of the car. 
\Resources 16 pixels to represent opponents and finish line; 15 pixels to represent the current car position in the race; 5 sound effects (two for the start, one for the engine sound whose pitch should increase with speed, one when the player loses the control of the car, one for passing the finish line), an input system (a key or a touch screen).
\LevelDesign 
Difficulty depends on the opponent behavior that should be controlled by basic heuristics (for example, the rubber band).
\Instructions Press the key (or tap) to increase speed, do nothing to slow down; racing fast for prolonged periods make the car uncontrollable and forces it to stop and restart the engine.  
\Concept{``.- .-.. .. ...- ."}
Players communicate with an astronaut (a non player-character) on a damaged spaceship, lost in space. The astronaut tries to contact someone using the only available mean of communication available, a binary signal, that is appears on the screen as a single pixel which can be on or off. The player in an operator in a space station that has a Morse code table to decipher incoming messages and a manual of the damaged spaceship with a map. The astronaut sends messages that the player can replay if needed. The player reply using morse code by pressing a key. Like in well-known text adventures, the player interact with the astronaut using basic verbs and composition of verbs and objects. The player can cancel a message using another key. The astronauts also sends short messages. The player will use the information in the manual to help the astronaut repair the ship to return to Earth. 
\Resources A white pixel, a button to send Morse code, a rewind button to replay messages or cancel an outgoing message, and the manual of the spaceship; sound effects should be added to improve engagement and immersion.
\LevelDesign This is a text adventure with a minimal and cumbersome mean of communication. Commands should be simplified to ease communication using the shortening used also in traditional text adventure in which "N" would correspond to "Go North" and "O" would correspond to "Open". It should be very short (because of the difficult communication interface) and mainly focused on creating empathy with the lost astronaut. 
\Instructions Help the astronaut come home. Listen to the Morse messages arriving from the damaged spaceship and use the manual to send instructions that will help the astronaut repair the damage and come home safe. 
 
\Concept{Simplified Fruit Ninja.}
A pixel is falling from the top of the screen. The player must click on the pixel before this reaches the bottom of the screen.
\Resources A falling pixel, a controller to click the pixel (mouse, touch, pad).
\LevelDesign In the first level, just one pixel appears at the top of the screen and it falls slowly. In the next levels, the falling speed and the number of pixels increases; additionally, there might be special pixels that players should not select. 
\Instructions Click on the falling pixels, before they reach the bottom.

\Concept{Minimal Wipe Out.}
Players must collect a sequence of pixels that fall down from the top of the screen toward the bottom. 
A pixel is collected when it reaches the bottom of the screen inside a target area, that is a small area in the middle delimited by using two pixels (one on the left and one on the right).
The player can move all the falling pixels on the screen at once, to the left or to the right, in order to make sure that the bottom one will fall into the target area;
in addition she can also slow down the falling speed of the pixels.
At each moment of the game there will always be five falling pixels on the screen, equally spaced on the y-axis of the screen, i.e., they all fall with the same speed. 
As soon as a falling pixel is collected, it disappears at the bottom of the screen and a new one will appear at the top.
The game ends when one of the falling pixel reaches the bottom of the screen outside the target area and thus it has been lost.
\Resources
The game requires the presence of 5 pixels that move from top to bottom and can be moved (all toghether) to the right or to the left.  
The player actions will require three buttons: one to move pixels to the right, one to move them to the left, and one to slow down their falling.
Finally, the target area at the bottom of the screen should be delimited using, at least, two fixed pixels.
\LevelDesign 
The Level Design is based on the following variables: (i) the initial speed of pixels, (ii) the changes over time of the falling speed, and (iii) the relative positions of the sequence of pixels (that appear will appear at the top of the screen).
The game allows both a procedural design of the levels or a deterministic design. 
In the latter case, players can exploit  their memory (to memorize the position of the pixel sequences) to improve their performance;
in this case, it is also possible to use color to make the sequence of pixels more easy to memorize.
\Instructions
Move the falling pixels on the screen toward left or right using the arrow keys in order to collect them inside the target area at the bottom of the screen.
You can also use the brake button to slow down the fall of the pixels.
Try to resists as long as you can without loosing any pixel! 
\Concept{Minimalist Shooting Hoops.}
The game is a minimalist wrap of a basketball shooting race. 
Players must keep a key pressed to load the shot and release it when the pixel lights up to mark the basket.
Players must score as many points as possible in a given time to move to the next level.
The start is announced by a single sound; a sound is played when players score; the end of the game is announced by a sound repeated three times. 
\Resources one pixel that can be on or off, one sound, one key or button to be pressed. 
\LevelDesign Level difficulty is based on overall time available to the player, the amount of baskets needed to move to the next level, the amount of time before the pixel lights up.
\Instructions a single sound marks the start of the game; hold the key to charge the shot and release it when the pixel lights up to shoot hoops; score as many points as possible before the time runs out; a cheering sound will accompany every scored point; three repeated sounds mark the end of the game.

\Concept{One Voice.}
The player, represented as a circle,  must dodge obstacles and enemies using her voice. Squares represent obstacles while diamonds represent enemies. The player can jump over enemies, climb obstacles or sneak under them. Points are collected for jumping over the enemies or sneaking under the obstacles. 
\Resources Geometric shapes for circle, square, diamond; a microphone and voice recognition software.
\LevelDesign This is an endless run so there is one level and difficulty increases over time. Initially, players will face an enemy and then an obstacles. Next, a sequence of enemies and obstacles will appear. Obstacles will form stairs that player can climb to higher positions making it easier for the player to jump over more enemies thus collecting more points. 
\Instructions Just say "up", to start the game, climb obstacles, sneak below them, and jump over enemies.
 
\Concept{Overtake}
When the output system turn on, the headlights of a car coming in the opposite direction; players have one second to change lane by using the input system (by pressing a button or a key); while holding the button, the players car will remain in the overtaking lane; but they will have a split second to get back to the main lane.
\Resources An input system (a button, better if a steering wheel that presses the "button" when turns counterclockwise over a certain point) and an output system (a pixel that turns on, a sound, a vibrating system).
\LevelDesign The frequency of incoming cars increases every time players complete a series of ten overtakes.
\Instructions You bet you can reach California by dawn; go up on your fireball and dart as fast as possible on the highway, put yourself on the overtaking lane to let the other cars eat dust; enjoy the breeze but beware that you little to re-enter the main lane.
 
\Concept{Pixelcraft Together.}
This is a multiplayer experience in which each player controls a pixel of a unique color within a persistent world. 
Players move their pixels using either touch (or a mouse); tap (or holding a mouse button) will draw a pixel on the world background. The world is a huge canvas where players can draw copies of their pixels. Will players communicate? How will they do it? What will they create? Will the world become a random collection of pixels or will players try to self-organize to create something meaningful? After a given number of actions have been performed or a given time has passed, the world is saved in a gallery. 
\Resources Multiplayer support with online storage, many colorful pixels, a controller (touch or mouse).
\LevelDesign There is no playable level, players will create it together. A cool down period between pixels' creation should be introduced. Optionally, obstacles and trails that players should follow could be introduced. cooldown between the creation of one's own pixel and another.
\Instructions Tap to move, hold to leave a footprint. 
\Concept{Purgatory}
Player move a white light (represented as a pixel) on a screen of 20x20 pixels, using a device equipped with accelerometers, and must captures blue lights (blue pixels) while avoiding red lights. 
\Resources sources of colored lights; a device with accelerometers.
\LevelDesign the first level contains just a blue light (one blue pixel) and a red one on a 20x20 pixels screen; the number of blue and red lights increase in the subsequent levels making it more difficult to reach the blue lights.
\Instructions you are an angel of purgatory who must recover all the souls that have managed to find peace; move and tilt your device to the white light to recover all the blue souls that are allowed to reach the heavens; avoid all the red souls that are not yet ready to do so.
 
\Concept{Quick Color.}
The screen displays a grid of pixels of different color; the grid colors change at fixed time intervals; 
the player must always select the only red pixel in the grid.
\Resources a grid of colored pixel, a target color, a pointer (touch or mouse).
\LevelDesign The first level consists of a 4x4 grid; colors change every 5 seconds; as the game progresses, the grid expands and the time intervals between color change is reduced.
\Instructions pick the red pixel before colors change. 

\Concept{Quick Reflexes.}
Two symbols appear on-screen: one is a directional arrow, the other one represents either the concept of equal or opposite. Players must press, within a certain time limit, the same arrow if associated with the concept of equal, the opposite one, otherwise. Players have a maximum number of mistakes they can do in each level, for example, by pressing the wrong arrow or no arrow at all. The game ends when that limit is reached. 
\Resources six symbols, four representing the directions, two for the concept of equal or opposite; a controller with four direction (a keyboard or a pad).
\LevelDesign a level consists of a number of configurations presented to players; level difficulty is based on the number of allowed errors, the time available to press the correct arrow, and the time before the next configuration appears. 
\Instructions Press the arrow in the same or opposite direction depending on whether the equale or opposite concept is shown. 
 
\Concept{Ringtone Master.}
The game screen consists of ten pixels that can be either black (when they are off), blue, or white (when they are on). Each pixel corresponds to one of 0-9 digits of the numeric pad. Pixels turn on rhythmically, and players must press the corresponding key on the numeric pad emitting a sound. Pixels will become blue moments before turning white to give players time to prepare. The series of sounds produced by following the rhythm imposed by the pixels make up a song. \Resources ten pixels, ten sounds, and a numeric keypad (keys 0-9). \LevelDesign a level is a sequence of pixels that turns on following a given rhythm, allowing players to play songs similar to those created with the composer software of Nokia3310. Level difficulty depends on the rhythm complexity, speed, and length of the melody. \Instructions Press the keys corresponding to the pixels that turned white and keep the rhythm. 
 
\Concept{Rog.}
Players are adventurer that must survive a linear dungeon they are exploring by fighting the enemies they meet. 
The game ends when players reach the dungeon exit. The current player view of the linear dungeon is represented by a character displayed on the screen. For example, an A might represent a door (the entrance of the dungeon); an I may represent a snake to defeat. Players might have an external legend to decipher the meaning of the characters or simply learn by trial and error without any additional resource.
Players have only one action (a key or a button to press). Combat works as follows. 
Players have only one statistics, the energy, that works both as a measure of attack force and life points. 
When players keep the key pressed, they consume energy, the longer they keep the key pressed the higher the energy they consume. Players' energy is represented by an ASCII extended character, with a value equal to the corresponding ASCII code (therefore between 0 and 255). The expended energy is shown on-screen as it is consumed.
For every encounter, a random number is generated (based on the type of enemy) and shown to the player. 
If the energy that the player has invested in the attack is lower than the random number, the enemy is defeated and the player can continue to the next encounter.
Otherwise, the player loses the invested energy and must attack again. 
When the players energy falls below 32 (the limit of visible ASCII characters), the adventurer dies. 
There might be elements in the dungeon that help the player (for example, a healing potion can restore energy).
This combat mechanics recreates the \textit{push-your-luck} experience of rogue-like games, pushing players to balance the amount of energy to invest in an attack, in the most efficient way possible, with a certain dose of luck to consider.
\Resources one ASCII character on-screen for the encounter; one for the energy invested in the attack; one for the current energy; one key.
\LevelDesign It follows the typical structure of rogue-like games presenting to players increasing powerful enemies and new elements such as traps, potions, and obstacles.
\Instructions Escape from the Dungeon; push the button to proceed; hold the button down to decide how much energy to use to attack enemies. If the energy expense is greater than the enemy attack, the enemy is defeated. If your energy drops to " ", you lost. 
 
\Concept{Role Pixel Game.}
In this minimalist role playing game, four pixels on the screen show the character's equipment (gray if common, blue if epic, orange if legendary, with associated a health value, damage, and percentage of increasing dodging), the health of the character (green, yellow, or red), the presence and level of the monster in the room (using the color coding used for the equipment), and finally the presence and type of treasure in the room (equipment or health potion Indicated by red color) or exit (white). Players can move to another room (using the four arrows to move in the corresponding cardinal direction), collect a treasures and potions, or attack. Players' score is based on the number of explored rooms, enemy killed, treasures, and equipment collected. 
\Resources four pixels and seven sounds for the steps or door that closes at the entrance in a room; the player's successful attack and dodging; the damage received successful attack; 
	collecting the room content, and victory. One input to hit and collect, four cardinal directions (arrows on a keyboard or swipe on a touch screen).
\LevelDesign The difficulty increases with an increasing number of enemy encounters, fewer potions, and a greater number of mandatory fights. 
\Instructions Explore the dungeon moving in the four cardinal directions to enter new rooms; press a key to attack and collect dungeon treasures, make sure you improve your equipment to face increasingly strong enemies and get healthy and except for exit. 
\Concept{See(D) Me Grow.}
Player plant a seed (a brown pixel) touching the screen. The seed blossoms over time (indicatively in half the pixel becomes green), 
	the longer the time the player spend watching the plant grows, the higher the probability that a rare flower will bloom (represented as a rare pixel color).
\Resources a pixel, two sounds (one for planting, one for blooming), a touch screen.
\LevelDesign Levels might involve increasing interaction with the seed to make it bloom or to avoid withering.
\Instructions Tap the screen to plant a seed, tap again to collect the blossomed flowers.
 
\Concept{Talking Robot.}
At the beginning, the computer emits a series of sounds representing all the letters of the alphabet in sequence from A to Z. 
Players must find a way to annotate the mapping between letters and sounds. 
Next, the computer emits a series of complete sentences, using the same sound-based coding like a sort of R2-D2. 
Players must decipher the sentences and annotate the solution (on a piece of paper or a notebook).
Finally, the computer show the solution (it could print it on screen or play it as audio) and players must check how well they deciphered the messages. There is no explicit score but just the players self evaluation of their own performance. 
\Resources there is not input just audio output; paper and pencil for the taking notes; a way to show the solution which could be player using the same audio or printed on screen. 
\LevelDesign the game starts with very short sentences that can increase in length when players move to the next level; reproduction speed might increase and the number of the sentence repetitions played may decrease to increase the difficulty.
\Instructions Listen to the 26 sounds representing the letters of the alphabet from A to Z and find a way to remember what sound is associated to each letter; next, a sentence will be played several times; decipher it  before it stops being repeated. 
 
\Concept{The Key}
A sequence of letters appears on screen, one at a time, and players must press the correct key as quickly as possible. \Resources a device with a keyboard to enter all the letters from A to Z and the digits from 0 to 9 (a computer or a mobile device with a software keyboard). \LevelDesign a level consists of a sequence of ten symbols. The first level asks players to press the key corresponding to the same symbol appearing on the screen. In the following levels, players may be asked to press the next or previous symbols of the one appearing on screen; the time available to press the key might also be reduced. \Instructions press the same symbol for the first level, and press the following/preceding symbol for the next ones. 

\Concept{They Never Stop.}
Players shoot advancing enemy pixels they are advancing. Enemies change their state periodically switching between an invincible state, a normal state and a brief vulnerable state.
Bullets recharge slowly and there are at most three live bullets at once. Thus, players must carefully decide when to shoot enemies to hit them in their vulnerable state and not wasting bullets when they are invincible. 
\Resources two pixels for player and the enemy, four colors, a key to press (or button). Sound and music to make it a complete rhythm game.
\LevelDesign one level lasts for around two minutes and the speed of enemies and their spawn rate increases when players step to the next level.
\Instructions Press the key to shoot the enemies when they are most vulnerable; do not waste bullets on invincible enemies.

\Concept{To Be Or Dot To Be.}
The game is a minimalist interpretation of the classic Frogger. A pixel is positioned at the lower edge of the screen. Players use a key to make the dot move of a fixed amount.
In the middle of the screen there is a row of 5 dots that moves horizontally: first from left to right and then from right to left. Players have to guide their pixel to the other side of the screen by crossing the line of moving pixels, without hitting the moving pixels.  
\Resources A pixel to use as the player character and for the moving line of pixels; 
two colors, one for the background and one for the dots;
a key.
\LevelDesign Each level requires players to guide five pixels across the screen. 
Initially, line of pixels moves slowly at a constant pace. Every time a pixel crosses the screen it is added to the moving line, becoming an obstacle for the next dot. So the second level will contain two lines of five dots each, the initial ones with the pixels that crossed in the first level. The two lines move in opposite directions.
\Instructions Bring the pixels across the screen without hitting the moving lines.

\section{Discussion}
\label{sec:discussion}
The game designs we included in the previous section provide a good overview of the variety of themes and mechanics of the concepts we received. 
There are games that are purely reactive (e.g., \textit{Dot-A-Mole} and \textit{Quick Reflexes}), involve chasing (e.g., \textit{Follow}), idle/waiting (e.g., \textit{See(D) Me Grow}), memory (e.g., \textit{Constellation}), quick reaction (e.g., \textit{The Key}, \textit{Quick Color}, and \textit{Quick Reflexes}), experiential (e.g., \textit{Pixelcraft Together}), and multiplayer (e.g., \textit{I am the Fastest!} and \textit{Pixelcraft Together}). Some mechanics are well-know (e.g., \textit{One Voice} has a rather typical mechanics with an experimental voice-based control); some are the same used in C64 and Apple II games (\textit{I Am Fastest} has the same mechanics of race track events on old Olympic video games). The combat mechanic of Rog recreates the \textit{push-your-luck} used in several tabletop games. \textit{Hop!} is inspired to the children jump rope games.
Several games are based on the players' memory (e.g., \textit{Talking Robots}, \textit{Constellation}), 
	others requires planning abilities (e.g., \textit{3.. 2.. 1.. Go!} and \textit{Kick the Engine}), 
	most of the are purely reactive. 
Note however that, some reactive games can be made memory-based through level design. For example, \textit{Circle Wave} could become memory based by making the sequence of circle waves deterministic which would be equivalent to asking players to learn the waves' behavior (similarly to what is done in Constellation).
There is also a wide variety of genres. There are racing games (e.g., \textit{Minimal Wipe Out}, \textit{Overtake}, and \textit{Kick The Engine}), rhythm games (e.g., \textit{Ringtone Master} and ), platformers (e.g., \textit{One Voice}), role-playing games (e.g., \textit{Rog} and \textit{Role Pixel Game}), casual games (e.g., \textit{Minimal Fruit Ninja}), adaptation of arcade games (e.g., \textit{They Never Stop} and \textit{To Be Or Dot To Be}), and experimental (e.g., \textit{Talking Robot}). Some games are completely abstract (e.g., \textit{Hop!}, \textit{Follow} and \textit{Constellation}), others are grounded in specific themes. 
It is interesting to note that, in terms of score system, most are win-loss games that challenge the player to complete as many level as possible, some have a scoring system, while others are completely experiential (e.g., \textit{Pixel With Friends} and \textit{See(D) Me Grow}).

If we analyze them using four criteria considered in \cite{DBLP:conf/fdg/NealenSB11} (system, control, visual, aural), we note that all games have minimalist systems with a very limited set of choices that may enable purely reactive behavior as well as planning (e.g., \textit{3.. 2.. 1.. Go!}) or deep gameplay using additional material in ``.- .-.. .. ...- ."). At control level, there is a neat distinction between games using keys or buttons and those needing a touch interface (which can in some cases replaced by a mouse pointer). Ten years ago, when minimalist game design movement started to get momentum \cite{DBLP:conf/fdg/NealenSB11}, touch interfaces were in their infancy and several minimalist games were initially design for the keyboard and then upgraded to touch \cite{TheImpossibleGame,Canabalt}. Today, touch interface has become a commodity and thus several designers view it as  minimalist. If we cross data of designers' age and the use of touch interface, we note that older designer tended to stick to the key and button approach whereas the touch interface was suggested mainly by younger designers.\footnote{We chose to leave the concepts anonymous and not to include any data about the age or other characteristics of the designers to avoid any sort of bias in the reader and for obvious privacy reasons.} 
Visually, all games are completely abstract. In aural terms, designers have used sound effects to reinforce the narrative (e.g., \textit{Kick the Engine} and \textit{Minimalist Shooting Hoops}); most designers used sounds for providing feedback while others made sound the main subject of the game (e.g., \textit{Talking Robot}).

The concepts highlight the important role of narrative even in minimalist games. Two games, \textit{Aliens Attack from Deep Space} and \textit{Minimalist Shooting Hoops}, at the system level, are \textit{identical} to the basic example games that were given in the invitation we sent to designers. The former asks the player to shoot the alien as soon as it becomes visible through the peephole (which is identical to the basic example game in Section~\ref{sec:games}); the latter asks the player to charge the shot by keeping the key pressed until the pixel turns on (which the same game with inverted control discussed in Section~\ref{sec:games}). However, the narratives that the designers have created around the same mechanics make the games feel completely different. Narrative is also used to create depth in a game with a cumbersome communication interface like ``.- .-.. .. ...- ." in which the use of a  printed manual may make it less minimalist but, at the same time, increase complexity, immersion, and depth.

\section{Conclusions}
\label{sec:conclusions}
We presented an  experiment on minimalist game design. We started from an example of the most straightforward game we could design (involving just one pixel and one interaction). We sent our minimalist game to students and professional game designers and challenged them to create original games using the least amount of visuals and user interaction. We did not impose any constraints;
we just provided our simple game as inspiration. We received more than 120 game concepts and presented some of them in this paper discussing the vision of Minimalist Game Design \cite{DBLP:conf/fdg/NealenSB11} that transpire from the concepts, ten years after the seminal paper by Nealen et al. \cite{DBLP:conf/fdg/NealenSB11} started the discourse on minimalism. We performed several analysis on the design documents we received to search for patterns and semantic structures. We applied a wide variety of well-known text mining techniques including word embeddings, topic modeling, and text clustering \cite{bojanowski2016enriching,joulin2016bag,joulin2016fasttext,grootendorst2020bertopic}. However, our analyses did not revealed interesting patterns; our results suggest that this is probably due to the relatively small number of concepts and the conciseness of the descriptions. We believe our experiment can be an interesting tool for teaching the essence of game design to students. Accordingly, we plan to propose the challenge again next semester to the incoming game design students. We consider this an ongoing experiment and we hope that increasing the amount of concepts will enable us to do more advanced analysis on the design documents in the future. We also welcome submissions of new minimalist design concepts from anyone interested in participating using the form at \url{https://forms.gle/uPk1aWkx9Gr9kVgk9}. 

\bibliographystyle{ACM-Reference-Format}

\end{document}